\documentclass{aastex631}

\begin{document}

\title{Semi-major Axis Jumps as the Activity Trigger in Centaurs and High-Perihelion Jupiter Family Comets}

\correspondingauthor{Eva Lilly}
\email{elilly@psi.edu}

\author[0000-0002-7696-0302]{Eva Lilly}
\affiliation{Planetary Science Institute 
1700 E. Fort Lowell, Suite 106 Tucson 
AZ 85719-2395, USA}

\author[0009-0002-9267-4661]{Peter Jev\v{c}\'ak}
\affiliation{Comenius University 
Mlynsk\'a dolina, 842 48 Bratislava, Slovakia}

\author[0000-0003-1800-8521]{Charles Schambeau}
\affiliation{University of Central Florida 
4000 Central Florida Blvd 
Orlando, FL 32816, USA}

\author[0000-0001-8736-236X]{Kat Volk}
\affiliation{Planetary Science Institute 
1700 E. Fort Lowell, Suite 106 Tucson 
AZ 85719-2395, USA}

\author[0000-0002-1717-2226]{Jordan Steckloff}
\affiliation{Planetary Science Institute 
1700 E. Fort Lowell, Suite 106 Tucson 
AZ 85719-2395, USA}

\author[0000-0001-7225-9271]{Henry Hsieh}
\affiliation{Planetary Science Institute 
1700 E. Fort Lowell, Suite 106 Tucson 
AZ 85719-2395, USA}
\affil{Institute of Astronomy and Astrophysics, Academia Sinica, P.O.\ Box 23-141, Taipei 10617, Taiwan}

\author[0000-0003-1156-9721]{Yanga R. Fernandez}
\affiliation{Department of Physics,
University of Central Florida,
4111 Libra Drive, \\
Physical Sciences Bldg. 430,
Orlando, FL 32816-2385}
\affiliation{Florida Space Institute,
University of Central Florida,
12354 Research Parkway, \\
Partnership 1, Orlando, FL, 32826, USA}

\author[0000-0001-9542-0953]{James Bauer}
\affiliation{University of Maryland, College Park, MD 20742-2421, USA}

\author[0000-0002-0439-9341]{Robert Weryk}
\affiliation{University of Western Ontario, 1151 Richmond Street, London ON N6A 3K7, Canada}

\author[0000-0002-1341-0952]{Richard J. Wainscoat}
\affiliation{University of Hawaii, 2680 Woodlawn Dr, Honolulu, HI 96822, USA}

\begin{abstract}

We present a dynamical study of 39 active Centaurs and 17 high-perihelion ($q > $ 4.5 au) JFCs with a focus on investigating recent orbital changes as potential triggers for comet-like activity. We have identified a common feature in the recent dynamical histories of all active Centaurs and JFC in our sample that is not present in the history of the majority of inactive population members: a sharp decrease in semi-major axis and eccentricity occurring within the last several hundred years prior to observed activity. We define these rapid orbital changes as ‘$a$-jumps’.  
Our results indicate that these orbital reshaping events lead to shorter orbital periods and subsequently greater average per-orbit heating of Centaur nuclei. We suggest the $a$-jumps could therefore be a major trigger of cometary activity on Centaurs and JFCs. Our results further imply that analyses of the recent dynamical histories could be used to identify objects that are currently active or may become active soon, where we have identified three such Centaurs with recent $a$-jumps that should be considered high-priority targets for observational monitoring to search for activity.

\end{abstract}


\section{Introduction}
\label{s.Intro} 

The Centaur population is comprised of small, likely icy bodies on dynamically unstable orbits transitioning from the trans-Neptunian region to the inner Solar System. Most Centaurs originate in the scattered disk, with a small fraction coming from other TNO populations and the Oort Cloud \citep[e.g.,][]{Gomes2008, Dones2015,DiSisto2020, Kaib2022}. 
Centaurs represent an interesting stage in the dynamical evolution of primordial small bodies scattered by planetary perturbations from the ``deep freeze'' of the TNO reservoir where they have been stored for billions of years. Centaurs enter the giant planet region as largely primordial objects and carry information about the composition of their parent populations closer to our doorstep. Once a Centaur crosses the orbit of Saturn, its dynamical evolution and thermal processing is rapid, driven by numerous close encounters with Saturn and Jupiter, where it eventually joins the Jupiter-family comet (JFC) population, collides with a planet, or is ejected from the inner Solar system \citep[e.g.][]{Tiscareno2003,Volk2008,Gkotsinas2022}. 

As the parent populations of Centaurs formed over a relatively large heliocentric distance range \citep[e.g.,][]{Dones2015,Morbidelli2020,Gladman2021}, one might
expect Centaurs to contain an abundance of various ices and exhibit comet-like activity even at
large distances, as has been observed on several long-period comets \citep{Meech2017,Bernardinelli2021}. Furthermore, if all Centaurs originated from the same parent population and followed similar dynamical routes to the giant planet region, we would expect to observe activity on all Centaurs far beyond the orbit of Saturn and the activity strength to mainly vary with heliocentric distance. We would also expect to observe transitions between regions where different activity drivers dominate, since the majority of Centaurs orbit outside the heliocentric region where H$_2$O sublimation plays a key role.

The low temperatures in the range of heliocentric distances where the active Centaurs have been observed suggests their activity is not driven by a water-ice sublimation, but either by sublimation of super-volatile species such as CO$_2$ (and CO if the Centaur has recently crossed over from trans-Neptunian region), or by crystallization of amorphous water-ice (AWI) \citep{Jewitt2009,Guilbert2012,Lilly2021a}. The surface temperatures are certainly high enough for Centaurs whose perihelia lie within the orbit of Saturn for some of the processes to yield vigorous gas production. However, only about 10\% of the population exhibits activity while the rest  does not, despite all occupying the same orbital space \citep{Jewitt2009,Chandler2020,Lilly2021a}, suggesting that Centaur activity depends on more than just the presence of volatiles and a favorable orbit. 

The differences we observe between individual bodies could be either due to different formation locations and therefore different internal compositions, or the amount of time spent in the Centaur region. \citet{Fernandez2018} have shown that the median lifetime of inactive Centaurs is about two times longer than that for active Centaurs, suggesting that activity is related to the residence time in the region. Several thermal modeling studies of dynamically evolved JFCs further suggested that the amount of time their parent object spends as a Centaur directly controls the level of thermal processing and the depth of pristine ice deposits available under their surfaces.  \citep{Fernandez2018,Guilbert2012,Gkotsinas2022}. However it remains unclear how exactly the residence time controls the potential for activity episodes in Centaurs, and how the active and inactive Centaurs differ from one another if their orbits are similar. Most of the dynamical studies concentrate on the Centaur population as a whole and investigate timescales in order of Myr and larger \citep[e.g.][]{Volk2008,Tiscareno2003,Fernandez2018}, when e.g., the typical lifetime of cometary activity driven by AWI on a Centaur is much shorter - it takes about 10$^4$ - 10$^5$ years until the thermal wave transforms available ice deposits in the nucleus and the activity ceases \citep{Guilbert2012}.

Here we present results of an analysis of the individual short-term dynamical histories of all 307 known Centaurs (as of December 2022) and 17 high-perihelion JFCs with the goal of identifying potential triggers of Centaur activity and thereby enable the identification of candidates expected to become active in the near future. 
We then utilize thermal modeling to investigate the thermal evolution of two end members of our active population sample: P/2019 LD$_2$ (ATLAS) and 167P/CINEOS (hereafter, LD2 and 167P, respectively). These two objects occupy opposite ends of the active Centaur population, with 167P being the active Centaur with the highest perihelion distance known to date, suggesting that it is among the dynamically newest active Centaurs, and LD2 actively transitioning into the JFC region, which is possibly near the end of its lifetime as a Centaur.

\section{Dynamical Integrations}
\label{s.dynamical_integrations} 

\subsection{Experimental design}
\label{ss.expdesign} 

Following \citet{Jewitt2009}, we define Centaurs as objects with perihelia between Jupiter and Neptune (i.e., $5.2~{\rm au}>q>30.0$~au) and semimajor axes of $a<30$~au, while excluding any Jupiter Trojans that happen to fall in these ranges. We also follow the methods of \cite{Schambeau2018DPS} in defining high-perihelion JFCs as those with $q\ge4.5$~au, because comets with such orbital parameters are likely to be the newest and least processed members of the group that have left the Centaur region only a few orbital periods ago \citep{Sarid2019,Guilbert2023}.
Using these definitions, our target sample includes all known Centaurs (including both active and inactive) and high-perihelion JFCs as of December 2022.

We created 50 dynamical clones for each target using multivariate normal distributions defined by orbital covariance matrices provided by the JPL Small Bodies Database\footnote{https://ssd.jpl.nasa.gov/sbdb.cgi}, in order to assess the level of divergence due to chaotic effects introduced by the object’s orbital element uncertainties as we track its orbital history.
Because Centaur orbits evolve in the giant planet region, they are strongly affected by dynamical chaos \citep[e.g.,][]{Tiscareno2003} and numerical integrations of a given object can typically only be considered reliable until a close encounter with a planet occurs, after which orbital clones offer only probabilistic indications of an object's long-term dynamical evolution. Over longer timescales, even these probabilistic indications cease to be meaningful, as the divergence of dynamical clones simply grows too large. Given these limitations, we only consider short integration times over which we can interpret our results with high confidence, namely where there are no close planetary encounters and the evolutionary paths of all clones for a given object are essentially identical. 

We conducted backward numerical integrations for the nominal orbit and all clone orbits of each Centaur and JFC in our sample for 200 years into the past using the Bulirsch-St$\ddot{o}$er integrator in the Mercury N-body integration package \citep{Chambers1997}, which has the ability to adapt timestep sizes when shorter timesteps may be needed to properly capture an object's dynamical behavior, e.g., during close planetary encounters. We use barycentric elements to display the final orbital evolution. To study the longer-term dynamical histories of our target objects and further investigate active Centaurs and JFCs with less prominent orbital changes, we also conducted backward integrations for 5000 years. 
All integrations accounted for gravitational perturbations from the Sun and the eight major planets and used an initial time step of 2 days for 200-year integrations, and a larger time step of 30 days for 5000-year integrations, respectively, in order to optimize CPU usage. 
Both of these timestep intervals are sufficient to track the  orbital evolution of Centaurs for which the gravitational influence of the inner terrestrial planets is negligible. Our initial test with shorter timesteps yielded essentially identical results as integrations using larger timesteps over the considered timescales. We have also tested the reversibility of our integration code on the case of LD2, which underwent an extreme orbit altering deep encounter with Jupiter (see left panels on Figure~\ref{fig.ajump_examples} by conducting forward integration from the end point of the backward integration at t -200 years. During such a close encounter the semimajor axis changes, which induces chaos.  While our integrations are not perfectly, fully reversible due to imperfect energy conservation and machine precision roundoff error, on the timescales of the reported $a$-jumps, our integrations are reversible to $\sim$1 part in 10$^9$ in semimajor axis and $\sim$1 part in 10$^7$ in mean anomaly (where mean anomaly will be the worst-conserved orbital element). 

Non-gravitational accelerations, such as those due to the Yarkovsky effect and asymmetric mass loss from cometary activity, were not included because their effect on JFCs and active Centaurs is expected to be minimal over the time span covered by our integrations. 

We performed numerical integrations as described above for all active and inactive Centaurs known as of December 2022 (39 and 268 objects respectively), and for 17 high-perihelion JFCs. The list of the active objects we analyzed is summarized in Table~\ref{tab.Orbparameters}. Upon completion of the integrations, we analyzed semimajor axis, eccentricity, and inclination evolution as functions of time and searched for large deviations from average orbital fluctuations. When such changes were found, we noted both the magnitudes of the changes as well as their durations.  For two selected cases --- 167P and LD2 --- integration output was used as input for our thermal model in order to characterize the effect of these bodies' orbital histories on the thermal processing of their nuclei (Section~\ref{s.Thermal_model}).

\subsection{Dynamical integration results}
\label{ss.Orb_changes} 

Our numerical integrations show that all active Centaurs and high-perihelion JFCs analyzed in our work underwent recent rapid orbit re-shaping events in the form of sudden decreases in semi-major axis distances and eccentricities that we refer to as $a$-jumps. The majority of $a$-jumps  occurred less than 250 years ago - only $\sim$12\% of the Centaurs we see active today underwent their most recent $a$-jump between 400 and 1000 years ago. The orbital changes were induced by close encounters with either Jupiter or Saturn, and have been extremely fast - on the order of several months in the most extreme cases, to several years. All noted $a$-jumps occurred before the divergence of the orbital clones when their orbital evolution was essentially identical. 

\begin{figure}[htbp]
    \centering
    \includegraphics[width=0.9\textwidth]{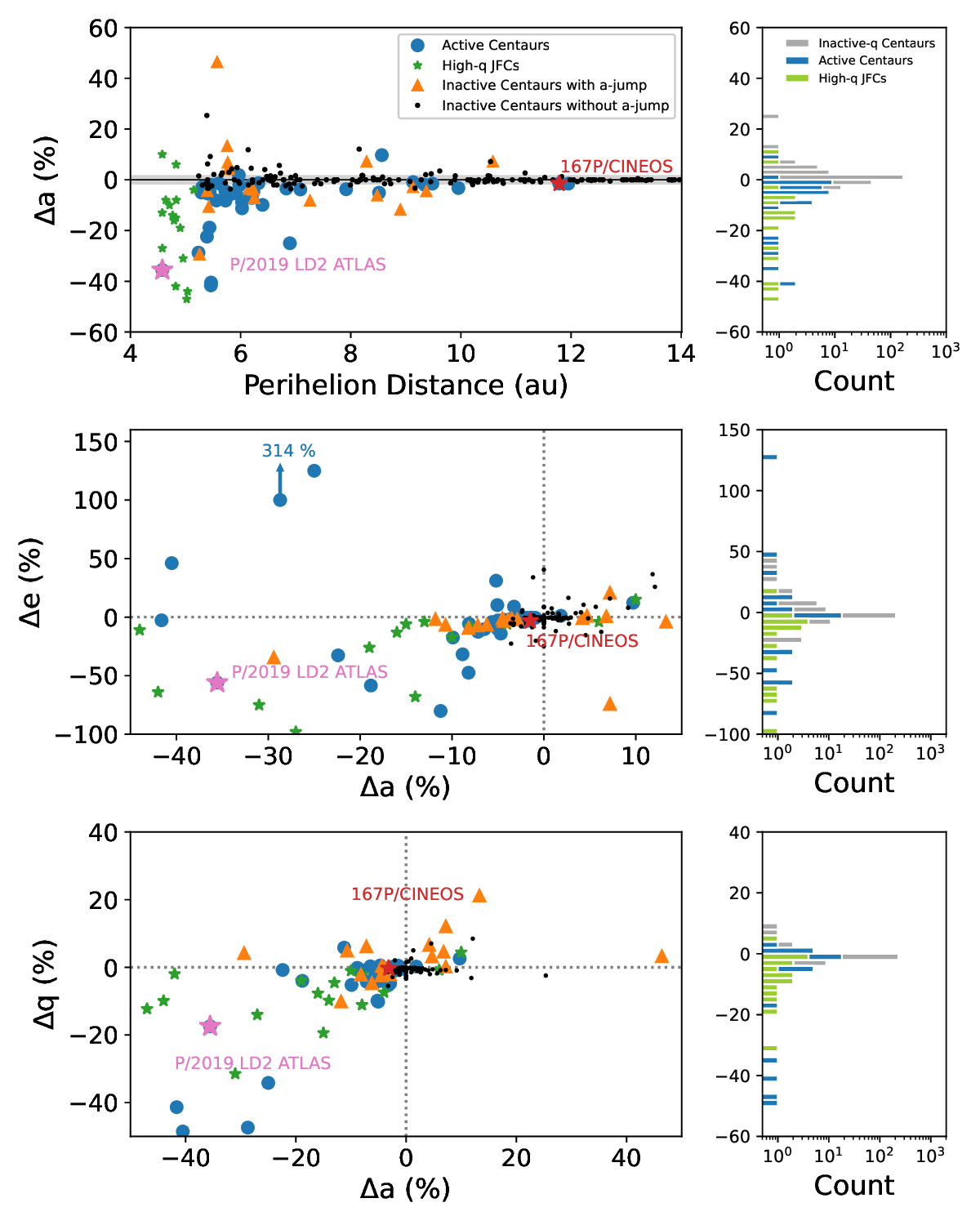}
    \caption{Left panels: (top) Percent change in semimajor axis from an object's most recent $a$-jump as a function of its post $a$-jump perihelion distance, where the grey horizontal region signifies the average amplitude of variations in $a$ for known inactive Centaurs; (middle) percent changes in eccentricity and semimajor axis; (bottom) percent changes in perihelion distance and semi-major axis. Right panels top to bottom: distribution of percent change in semimajor axis, eccentricity and perihelion distance for active and inactive Centaurs and high-perihelion JFCs. All depicted changes were fueled by the object's most recent $a$-jump in case of active bodies, or the highest amplitude change in the given element over the inspected time period.}
    \label{fig.da_vs_q}
\end{figure}

\begin{figure}[htbp] 
\centering
    \includegraphics[width=\textwidth]{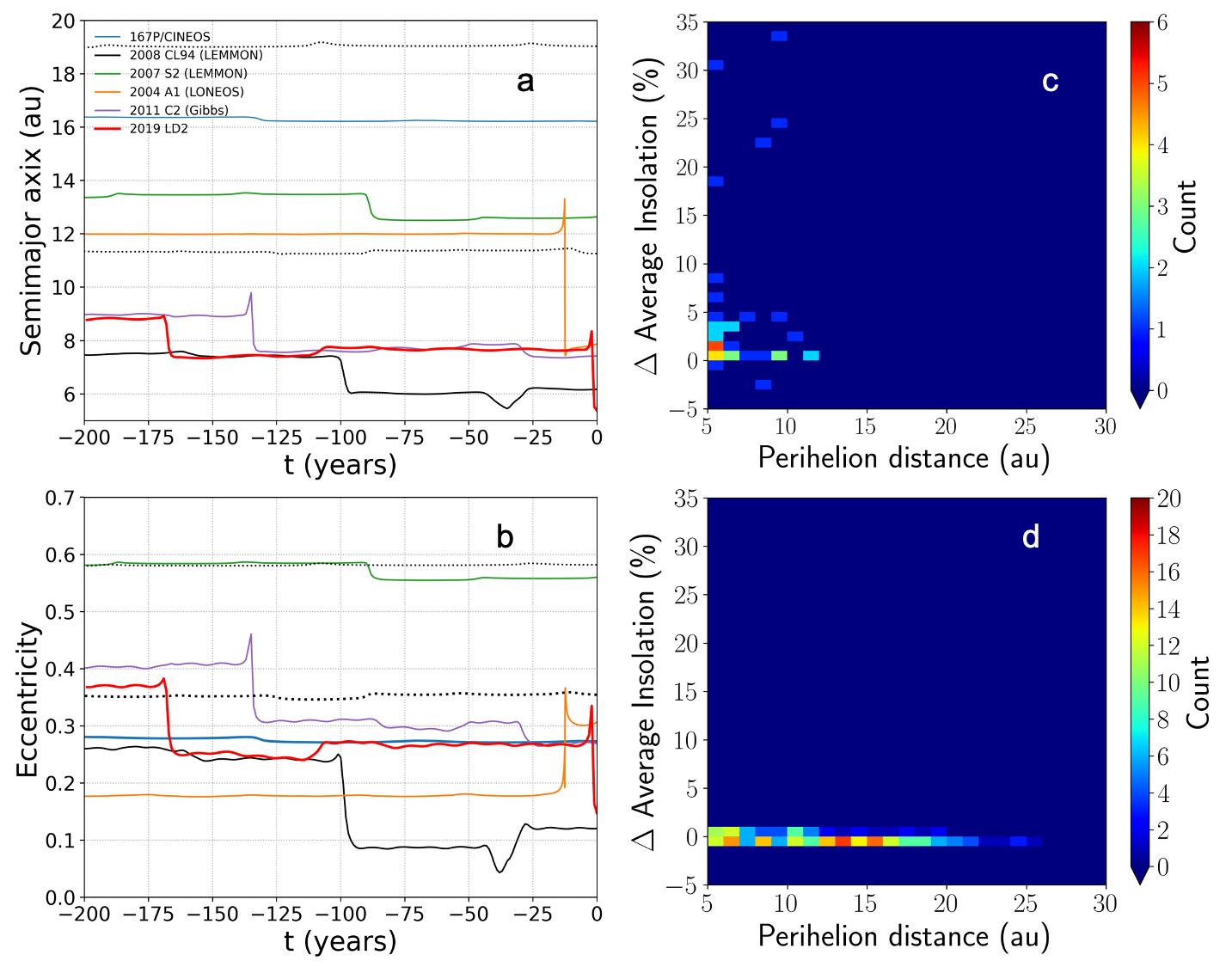}
    \caption{Panel a: most recent $a$-jumps - changes in semi-major axis  accompanied by changes in eccentricity (panel b) evident in the orbital evolution of selected known active Centaurs compared to a typical recent evolution of two inactive Centaurs (2013 JX$_{14}$ and 2003 QD$_{112}$) without the $a$-jump feature (black dotted lines). Some objects can undergo several $a$-jumps in a row. Time is relative to the epoch of orbit for integration. Only nominal orbits are shown for clarity. Panel c: percent change in the average per-orbit insolation of active Centaurs caused by the object's most recent $a$-jump as a function of its pre $a$-jump perihelion distance.  Panel d: percent changes in the average per-orbit insolation of inactive Centaurs caused by the object's largest varation in semi-major axis and eccentricity as a function of its pre $a$-jump perihelion distance
    } 
    \label{fig.ajump_examples}
\end{figure}

The changes in semi-major axis, $\Delta a$ was larger than $\sim$0.5~au in 74.5\% of the active Centaurs and JFCs we investigated (Table~\ref{tab.Orbparameters}). The most extreme $a$-jump occurred in the case of P/2004 A1 (LONEOS), which is now designated as 450P/LONEOS, where the Centaur's semi-major axis decreased by 5.2 au (more than 40\% of the original value of 12.7 au) over several months after a close encounter with Saturn (encounter distance of $<$~0.03 au) in 1992. The remaining Centaurs underwent milder $a$-jumps by about 0.2-0.4 au inwards. We hypothesize that these gentler orbital changes might be sufficient to trigger activity if they occur for objects with higher perihelia that are likely dynamically new in the region. The top left panel of Figure~\ref{fig.da_vs_q} suggests the dependence of the a-jump severity necessary to trigger activity on object's perihelion distance.

The change in semi-major axis in active Centaurs is typically accompanied by a change in eccentricity of similar magnitude, as seen in Figure~\ref{fig.da_vs_q}, which shows parameters and the distributions of the element changes for the most recent $a$-jumps for the objects we investigated.  The majority of active Centaurs also had corresponding changes in perihelion distance and inclination, which were typically very small -- less than 1\% of their original values. Only a handful of Centaurs, including 450P and LD2, with most extreme $a$-jumps ($\Delta a>\sim$10\%) are exceptions to this rule. The median value of $\Delta a_
{Cen}$ = -0.62 AU (or -5\% of the pre- $a$-jump value) and $\Delta q_{Cen}$ = - 0.08 AU (or -0.8\% of the pre- a-jump value). JFCs generally appear to undergo more pronounced changes in both elements than active Centaurs, with $\Delta a_
{JFC}$ = -1.2 AU (or -13.9\% of original value) and $\Delta q_{JFC}$ = -0.4 AU (or -7.4\% of the pre- a-jump value). In this way $a$-jumps cause sudden orbital re-shaping from eccentric to more circular  forcing the Centaur to experience more frequent perihelion passages than it would have before the orbital change, and most importantly to receive higher mean energy flux per orbit, which translates into higher mean per-orbit temperature and/or peak temperature \citep{Prialnik2009,Gkotsinas2023}. 
Figure~\ref{fig.da_vs_q} shows a single case - Centaur 166P/Neat where the semi-major axis in fact increased, but further examination shows that this increase was a result of an earlier $a$-jump after which the object in question spent several decades on orbits with smaller $a$, followed by additional close encounters with Jupiter that increased $a$ to the values we see today.  This is yet another example of the rapid and complex orbital evolution of objects on the edge of Jupiter's orbit, that can switch back between JFC and Centaur region in a matter of several orbital periods.

Remarkably, $a$-jumps are not present in the recent orbital histories of the vast majority of known inactive Centaurs over the timescales that we considered here. Our numerical integrations of 268 inactive Centaurs show that they typically undergo limited changes in $a$ and $q$ with a median maximum absolute change of 0.5$\pm$1.0\% in semi-major axis over our integration periods. KS-test comparing the maximum amplitudes in the variations of semi-major axis, eccentricity and perihelion over the inspected time period between the active and inactive Centaur populations suggests we can reject the null hypothesis that these two samples were drawn from the same populations at 95\% confidence level, meaning the two groups underwent different dynamical evolution paths. Left panels (a,b) on Figure~\ref{fig.ajump_examples} shows some of the prominent $a$-jump cases we have identified compared to the orbital evolution of two inactive Centaurs without identified $a$-jumps in their dynamical histories: 2013 JX$_{14}$ and 2003 QD$_{112}$. Note that all our plots use barycentric orbital elements to avoid sinusoidal patterns due to Jupiter's orbital period and the related movement of the Sun around the Jupiter-Sun barycenter \citep[e.g.,][]{Gladman2021} which do not produce meaningful differences in the thermal evolution of distant objects.

\section{Thermal Modeling}
\label{s.Thermal_model}

To assess the thermal effects of the orbital re-shaping events we noted in our dynamical integrations, we utilize a simple thermal model including heat diffusion to simulate the surface and interior heating received by a Centaur during an $a$-jump.
Similar approaches have been applied to describe the thermal evolution of Centaur interiors (e.g., \cite{Sarid2019, Gkotsinas2022, Guilbert2023}).
For simplicity we aimed to estimate the upper limit on the heating received by a Centaur during an $a$-jump using a spherical nucleus heated on the sub-solar point (i.e., a point always directly facing the Sun).
Note that an a-jump will presumably raise the temperature off of that point as well, and other factors, such as rotation rate and object's shape and obliquity will have implications for the effectiveness of an a-jump, i.e., even small a-jumps might produce activity if an object has slow rotation and/or large obliquity (detailed investigation of these effects can be found in \citet{Guilbert2012}). 
Spacecraft visitations to a variety of comets and the TNO (486958) Arrokoth have shown indeed how diverse and non-spheroidal shapes small bodies can have. 
However, such in-depth investigation including modeling the shape and spin is beyond the scope of this paper.

For the thermal modeling performed for this investigation, we seek to understand the bulk thermal impulse received by a nucleus as a result of an $a$-jump and the extent to which these orbital evolution events could trigger activity in a previously inactive body. 
We assume that some of our considered objects have probably experienced episodes of activity in the past but were temporarily inactive prior to their most recent $a$-jump (and the initiation of our thermal model) due to the depletion of accessible near-surface volatiles which occurred during their last active episode.

The thermal model solves the one-dimensional (1-D) heat equation through a porous matrix using a Crank-Nicolson method to solve an intrinsically non-linear equation using a 1-D grid.
A 1-D approach is sufficient for our investigation due to the relatively large sizes of the bodies being studied where we expect the vertical heat transport to dominate over the lateral heat transport due to the low thermal inertia of the nuclei materials \citep{Prialnik2004, Groussin2019}. 
The heat equation, solving for the temperature profile ($z$ is depth below the surface), is given by

\begin{equation}
\label{eq:heat_equation}
\frac{\partial T}{\partial t} = \alpha \frac{\partial^2 T}{\partial z^2}
\end{equation}

\noindent where $\alpha = k/(c_p \rho)$ is the thermal diffusivity, $k$ is the thermal conductivity, $c_p$ is the heat capacity, $\rho$ is the density of each layer.
For these material thermal properties we used suggested values from \citep{Groussin2019} based on recent space-craft visited cometary nuclei: $k$ = 0.006 W m$^{-1}$ K$^{-1}$, $\rho$ = 532 kg m$^{-3}$, $c_p$ = 770 J kg$^{-1}$, K$^{-1}$. 
Using these value results in a value of $\alpha$ = 1.46$\times$10$^{-8}$ m$^2$ s$^{-1}$ and a thermal inertia $I = \sqrt{k \rho c_p}$ = 50 J K$^{-1}$ m$^{-2}$ s$^{-1/2}$.
The surface boundary condition is given by equating energy balance at the surface of the nucleus,

\begin{equation}
(1-A) \frac{L_{\odot} }{R_H^2} = \epsilon \sigma T^4 + k \frac{dT}{dz}
\end{equation}

\noindent where solar energy input, blackbody emission of the nucleus, and heat conduction into the subsurface are included. Variables in the equation are as follows: the bond albedo ($A$), solar constant ($L_{\odot}$), heliocentric distance ($R_H$), emissivity ($\epsilon$).
Since the surface properties of most Centaurs are not well constrained, for our modeling efforts, we chose ensemble values for known active small bodies where we assume the surface is consistent with being ice-less and covered with a dust mantle: geometric albedo of 4\% (and thus a derived bond albedo of $A = 0.012$), and emissivity of 0.95 \citep{knight2023}.
The heating received by the spherical nucleus in our model is controlled by the surface boundary condition which is updated during the simulation using the object's heliocentric distance throughout its orbit provided by our orbital integrations.
For the thermal modeling included in our investigation we are using a time step of 1 day.

Our thermal modeling efforts start with choosing an initial structural layering and temperature profile for a progenitor nucleus's interior radial profile. 
We begin with a progenitor nucleus with an initial homogeneous interior at a temperature of 5 K with the material thermal properties fixed at the values described earlier.
We do not include phase transition of volatile species into our simple model here in order to minimize the complexity of this initial study and are here focusing on a preliminary assessment of the potential for the likelihood of interior temperature changes inducing periods of new activity. 
A future more detailed study will be undertaken to assess more precisely how much supervolatile sublimation is ongoing or the extent of AWI crystallization experienced during an $a$-jump for individual objects.

After the creation of a progenitor nucleus, we then simulate its thermal evolution during the time that it resides in the TNO region prior to becoming a Centaur.
To do so, we assume an initial circular orbit with a semimajor axis of 40 au for the progenitor nucleus, and run our thermal model until the radial temperature profile has reached a steady state.
Next, we change the object's orbit to an elliptical orbit to simulate a simplified inbound journey from the TNO region to a position in an individual object's past orbital history. 
Determining this elliptical orbit requires specifying a heliocentric distance from the object's previous orbital history (here is label as $R_{\circ}$) in which to insert the progenitor nucleus.
We use the dynamical integrations described in Section \ref{s.dynamical_integrations} to determine $R_{\circ}$ -  the heliocentric distance at a point of time in the past when the orbital clones of an object's backwards orbit integrations have not diverged significantly, and its orbit is well constrained. For the preliminary thermal modeling included here we choose this epoch by visual inspection of the orbital evolution plots. It is likely most Centaurs with lower perihelia might have experienced multiple close encounters with either Saturn or Jupiter resulting in orbital changes, however as our knowledge of these changes is probabilistic only, it is not in the scope of this paper to model the thermal evolution prior the most recent $a$-jump. We can merely assume our test bodies have undergone some several thermal processing cycles typical for JFCs before \citep{Gkotsinas2022}, thus here we present a lower limit on the possible thermal processing of the inspected objects.
Once this $R_{\circ}$ has been identified, an elliptical orbit is determined by finding the semi-major axis and eccentricity of an orbit with aphelion distance of 40 au and perihelion distance of $R_{\circ}$. Once at $R_{\circ}$, we continue modeling each object's thermal evolution following its known previous orbit, where the heliocentric distances used for the thermal model's surface boundary condition are from the past orbit integrations. 

\begin{figure}[htbp]
\centering
    \includegraphics[width=0.8\textwidth]{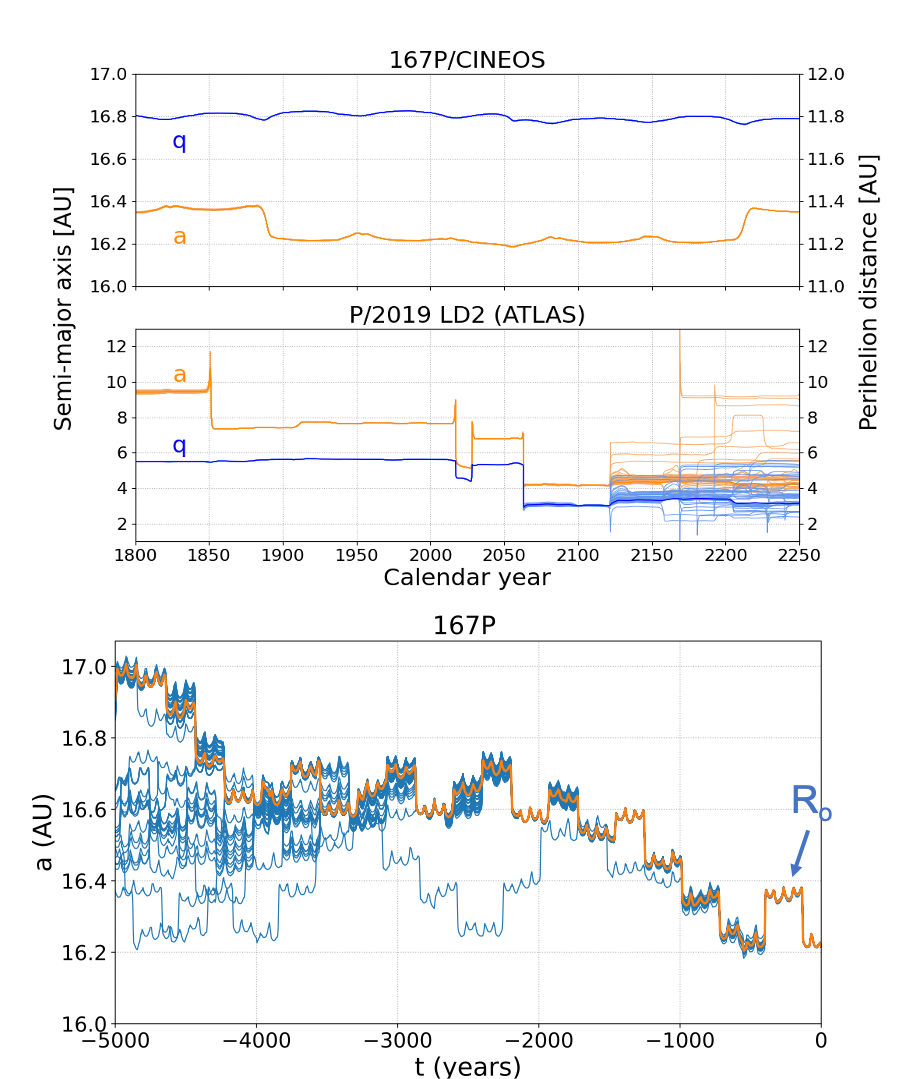}
    \caption{Orbital history of two active Centaurs at the opposite edges of Centaur region - 167P (Top panel) and LD2 (Bottom panel) The left y-axis shows the semi-major axis, the right y-axis shows perihelion distance Both nominal orbit and clones are depicted with lines of the same color. LD2 has experienced multiple close encounters with Jupiter leading to substantial orbital changes and is an example of a Centaur with prominent $a$-jumps. Such close encounters introduce chaos leading to clone divergence and difficulty in assessing the orbital parameters of LD2 before $\sim$1800 and after $\sim$2125. Bottom panel: The dynamical evolution of 167P in the past 5000 year for the nominal orbit (orange) and 50 orbital clones (blue). The clones of 167P follow essentially identical evolution down to 1800 years ago when they diverge due to wider encounter with Saturn after which the semi-major axis started to drift inwards. 167P is an example of a `drifter' Centaur. The arrow denotes part of the orbital history where the heliocentric distance R$_o$ is determined.}
    %\note{update figure, y-axis on bottom panel}.}
    \label{fig.167P-LD2-ajump}
\end{figure}

For illustration purposes, we apply the thermal model described here to two specific Centaurs, 167P and LD2 (see Section~\ref{s.Intro}), with a focus on assessing how differences in pre- and post-$a$-jump orbits affect the thermal conditions in the nuclei, which could influence their comae gas and dust productions.
We focus on these two objects because their current orbits place them at opposite ends of the heliocentric distance range over which active Centaurs are found and thus they experience the two extremes of heating environments possible for Centaurs. 
According to our integrations, 167P is a fairly new arrival to the $r_h<14$~au zone, and therefore may have experienced less thermal processing and volatile depletion compared to a typical JFC \citep{Gkotsinas2022}. 
In contrast, due to a set of close encounters with Jupiter in the last 300 years \citep[e.g.,][]{Kareta:2020,Steckloff:2020,Hsieh2021}, it is not possible to ascertain the exact orbital history of LD2 beyond $\sim$300 years ago, but as a recent escapee from the Centaur``Gateway" region, it has likely experienced more complex thermal processing \citep{Sarid2019, Guilbert2023}. 
Finally, there are good observational records available covering the active behavior of both bodies for several years after their discoveries. 
The epoch of $R_{\circ}$ for LD2 has been chosen at the year 1900 and is set at the beginning of x-axis the top panel of Figure~\ref{fig.167P-LD2-ajump}, $R_{\circ}$ epoch for 167P is annotated by an arrow on the bottom panel of the same figure.

\section{Case Studies}
\label{s.case_studies} 

\subsection{167P/CINEOS}

167P has a radius of $r\sim33$~km \citep{Bauer2013}, making it a medium sized Centaur, and is one of the two most distant known to be active: with $q=11.784$~au, it was observed to be active at a heliocentric distance of $r_h=12.23$~au \citep{Jewitt2009}, which is close to the outer boundary of the AWI crystallization zone \citep{Guilbert2012}. It was discovered on 2004 August 10 by the Campo Imperatore Near Earth Objects Survey (CINEOS) and appeared asteroidal at the time \citep{Boattini2007}, but a year later, it was reported to have developed an asymmetric coma\footnote{IAU Circ. 8545: C/2004 PY$_{42}$; C/2005 K$_{2}$, 2005 June 17}. This object underwent its most recent $a$-jump 139 years (or about 2 orbital periods) ago due to a distant $\sim$ 4 Hill radius encounter with Saturn, as seen on Figure~\ref{fig.167P-LD2-ajump}. This $a$-jump was one of the mildest found in our integrations for active Centaurs: the semi-major axis decreased by only 0.24 au (less than 1.5\% of the original value), but long-term numerical integrations show 167P has been slowly drifting inwards: its semi-major axis has been slowly decreasing for the past $\sim$2000 years from 17~au to 16.2~au, where the $a$-jump we identify occurred at the very end of this drift period. The perihelion distance of 167P gently has only oscillated around a value 11.8~au with no major changes. The object's orbital clones diverge due to chaos prior to this point in time, making the the object's orbital history prior to $\sim$ 2000 impossible to determine with much certainty. In about 200 years the semi-major axis will shift again to the pre-jump value, possibly quenching the activity. The $a$-jump we study here was the latest sudden reshaping of the orbit, and could have been the final impulse to heat the nucleus just enough to enable the onset of activity. 

The left set of panels of Figure~\ref{fig.LD2-167P-temp} shows the results from our thermal model for 167P: the surface and radial profile temperature of 167P during three different perihelion passages (in 1804 before its 1884 $a$-jump, in 1938, and during its last perihelion passage in 2003 April). The temperature is seen to increase by no more than a couple of degrees both at the surface and in the interior to a depth of $\sim$ 10 meters. Below this depth, the radial profiles for each epoch of perihelion passage converge to a temperature of $\sim$ 85 K. Interestingly, the hottest temperatures obtained for 167P through the modeling are too cold for AWI crystallization driven activity \citep{Prialnik2004}, but are too hot for the survival of sub-surface CO or CO$_2$ ice deposits to be present to drive activity through sublimation ($\sim$ 50 K and $\sim$ 80 K for CO and CO$_2$, respectively). The thermal modeling does show that the thermal wave penetrates deeper after the $a$-jump, however at the sub-solar point used for the simple thermal model included here the sub-surface region below this point is depleted of both CO and CO$_2$ pure ice species. This is not an unexpected result given the recent works in the literature focused on more detailed analyses of TNOs and Centaur interiors and the survival of pure super-volatile species over the age of the Solar System \citep{davidsson-2021thermal-model, lisse2022, parhil-prialnik-2023}. A possible explanation of 167P's observed activity could be the presence of sub-surface CO$_2$ ice deposits at higher effective latitudes away from the sub-solar point being activated by the deeper penetration of a thermal wave at those locations. 

In 167P's case, the $a$-jump may have caused temperatures to spike just enough to set off either sublimation of sub-surface CO$_2$ deposits, or to trigger AWI crystallization in layers that were previously intact, as discussed in \citet{Guilbert2012}. Given the time frame over which we modeled the object's temperature evolution, and its past dynamical evolution at high perihelion distances bracketed by clone evolution, we can rule out H$_2$O-sublimation driven activity.  The work of \citet{Guilbert2012} argues that at such large heliocentric distances (167P was at $r_h\sim12.3$~au when discovered), AWI crystallization should be minimal, and beyond $\sim12$~au, this activity driver can be ruled out for 167P as well. However, there is a possibility that it could have been weakly active even before the $a$-jump, with activity sustained by CO$_2$ sublimation, since 167P is very likely a dynamically new, ice-rich Centaur and the temperatures of the nucleus from our model were high enough (just above 80K) to allow its sublimation. Also \citep{Davidsson2021} have also shown that the process of CO segregation, where supervolatiles like CO and CH$4$ trapped and preserved in CO$2$ may be released during sublimation of CO$2$ at the front, or from even larger depths, could dominate beyond $\sim$10~au where crystallization rates drop. Interestingly, recent James Web Space Telescope (JWST) observations indeed confirm the presence of CO$_2$ ices on surfaces of other Centaurs at similar heliocentric distances \citep{Licandro2023} providing observational evidence that the volatile species necessary to produce comae activity are present in this region. However, the identification of 167P's activity driver is beyond the scope of this manuscript and indeed may not be identifiable until gas comae measurements are made of this body in the distant future.

\subsection{P/2019 LD2 (ATLAS)}

LD2 is a small outer solar system object with a radius of $r\lesssim1.2$~km \citep{Schambeau2020, Bolin2021} that is currently undergoing a much anticipated transition to a JFC via the JFC Gateway \citep{Steckloff:2020}, orbiting just inside the Centaur region border with $q=4.578$~au. It is therefore outside our formal definition of a Centaur, but given the fact that it experienced an $a$-jump that carried it inside Jupiter's orbit only two years ago, it is a perfect example of the dramatic orbital changes, and therefore intense thermal environment changes, that $a$-jumps can produce.  LD2 was discovered in 2019, but has exhibited activity in precovery images dating back to 2018 \citep{Schambeau2020}. Our integration results, which are in line with \citet{Steckloff:2020}, show that LD2 has experienced two rapid $a$-jumps.  The first one occurred $\sim$175 years ago, lowering the semi-major axis from $\sim9$ to 7.5~au. The most recent $a$-jump occurred in February 2017 and pushed the semi-major axis down to 5.3 au within the next six months (see Figure~\ref{fig.167P-LD2-ajump}). LD2 is also one of the very few active Centaurs where the close encounter led also to a major change in perihelion distance - by almost 1~au. As LD2 was not detected near its expected position in archival images from 2017 March, we can assume it was not active at the time, suggesting the onset of activity on LD2 was relatively sudden and took place between mid-2017 and early 2018 \citep{Schambeau2020}. Its orbital clones start to diverge $\sim400$ years ago when it experienced another close encounter with Jupiter, limiting analysis of its dynamical history prior to that point to statistical assessments only.  We can only infer this is likely not the first significant heating episode for LD2 given the proximity to Jupiter: the object's perihelion distance was between 4~au and 6~au in the past 5000 years with its semimajor axis mostly constrained between 4~au and 12 au, and future $a$-jump around 2060 will push both its semi-major axis and perihelion distance towards JFC-like values.

\begin{figure}[ht]
    \centering
    
    \includegraphics[width=\textwidth]{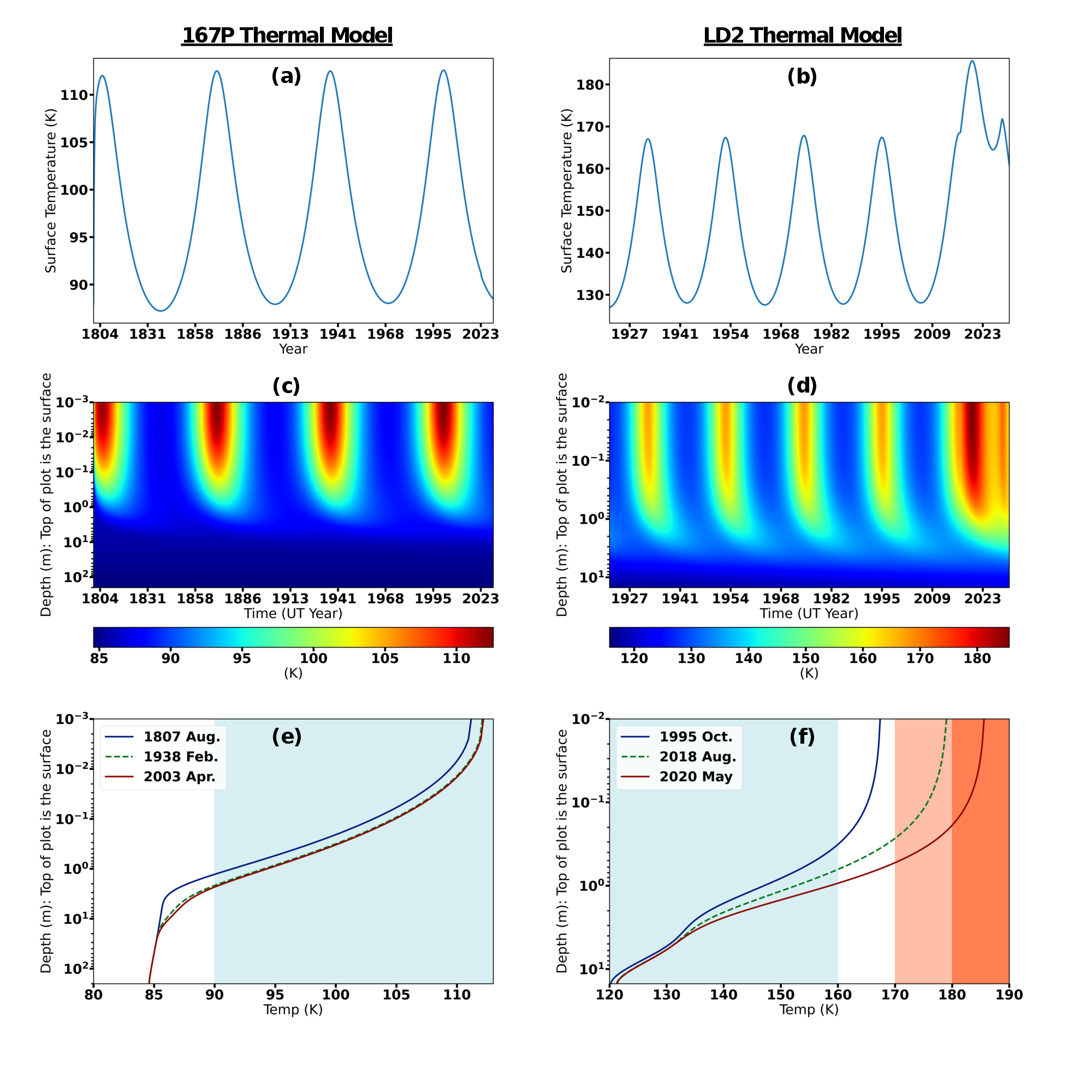}
    
    \caption{Panels a: surface temperature evolution, c: evolution of interior's temperature profile, and e: interior radial's profile for temperature during three different perihelion passages of 167P - just before its 1898 $a$-jump, in 1938, and during its last perihelion passage in 2003 Apr. before its 2004 Aug. discovery while active. The light-blue shaded region displays the AWI crystallization temperature regime ($\sim$ 90 - 160K) containing all three temperature profiles for 167P. Right-column, panels b - f: same quantities as in left-column pictured for LD2. A more pronounced temperature change can be seen for LD2 due to the more significant $a$-jump (top-right). The temperatures reached for LD2's interior are higher due to its orbit being closer to and now interior to Jupiter's (middle-right). The shaded regions on panel f indicate thermal regimes for different proposed activity drivers: AWI crystallization between 90 - 160 K (light blue), beginnings of water ice sublimation at $\sim$ 170 K (coral) and beginnings of vigorous water ice sublimation at $\sim$ 180 K (light red). The temperature evolution displayed pre- and post-$a$-jump may display a transition between activity driving mechanisms in play for LD2.}
    \label{fig.LD2-167P-temp}
\end{figure}

The right panels of Figure~\ref{fig.LD2-167P-temp} display the results of the thermal modeling of its most recent $a$-jump (Section~\ref{s.Thermal_model}), specifically the evolution of the radial temperature profile at the sub-solar point over the past $\sim100$ years until the year 2030. As can be seen from the figure, a thermal wave penetrates deeper into the interior of LD2's nucleus due to the $a$-jump, increasing the temperature by more than 20K, almost an order of magnitude larger change than in the case of 167P due to LD2's more dramatic orbital re-shaping and also perihelion decrease. We discuss the thermal implications of both the orbit re-shaping and a perihelion decrease in Section~\ref{ss.T-implications}. Over the past $\sim$ 175 years, LD2 was in a mostly stable orbit where a repeated cycles of interior heating and cooling occurred, as seen by the repeated patterns to the left of Figure \ref{fig.LD2-167P-temp}'s top-panel. The period following LD2's first $a$-jump 175 years ago likely depleted its near-surface volatiles, leaving the nucleus largely inactive for most of the last century. The following close encounter with Jupiter in February 2017 caused the inward journey for LD2 by means of the $a$-jump and forced the interior heat wave to propagate deeper, triggering phase transitions in the volatiles that have been demonstrated as plausible drivers of the cometary activity we observe today.
 
Figure~\ref{fig.LD2-167P-temp} (right-column panels) shows the evolution of surface temperatures, the propagation of heat into the interior, and the interior temperature profiles for three more recent peak periods of surface heating before, during and after the most recent $a$-jumps. The intense increase in both surface and interior temperatures is evident, explaining the onset of vigorous activity. Figure \ref{fig.LD2-167P-temp} (bottom-right) displays the interior's radial temperature profile for the peak surface temperatures received by LD2 during its pre-$a$-jump last perihelion passage in 1995 October (blue curve), the peak temperature achieved during 2018 August when the first observational evidence for activity was observed \citep[green dashed curve;][]{Schambeau2020} and finally for the peak surface insolation received by LD2 post-$a$-jump (red curve).  The transition of LD2's interior's peak temperature from being too cold for efficient water-ice sublimation to being hot enough for vigorous water ice sublimation provides a plausible explanation for the active behavior displayed shortly after its discovery during its most recent apparition \citep{Kareta2021,Bolin2021}. LD2's near-surface water-ice reservoir may have undergone a period of vigorous sublimation, while in the deeper interior (layers between 35 - 100 cm), AWI could also still potentially be present and could be heated sufficiently to contribute to observed outgassing due to crystallization.

\section{Discussion}
\label{s.Discussion} 

\subsection{Time Lag Between $a$-jump and the Activity Onset}
\label{ss.lag-implications} 

 Our numerical integrations show there is usually a time lag of varying length between the most recent $a$-jump and observed activity onset on Centaurs. In several notable cases of strong $a$-jumps, such as in the cases of 450P/LONEOS, LD2, C/2020 Q2(PANSTARRS) and others, activity onsets were almost instantaneous, e.g., within a year, while for other objects, lags between $a$-jumps and observed activity could be several decades to hundreds of years long, with the average value of 109.6$\pm$129.9 years. It appears this time lag is a real feature considering several Centaurs with deep prediscovery images still show point-source like nuclei taken months after $a$-jumps \citep[e.g. 167P, LD2, 2013 UL$_{10}$;][]{Mazzotta2018,Sarid2019}, but before the objects in question were flagged as active bodies. 
 
In principle, for well documented cases when activity onset can be identified with reasonable accuracy and precision (e.g., within a few years), the length of a time lag could help to identify the volatiles and processes driving the gas and dust loss and the inner structure of the Centaur nuclei. This is because the length of the lag should depend on the time needed for a thermal wave to travel through nonvolatile surface material before reaching subsurface ice, thus setting off the phase transition of those ices \citep{Gkotsinas2022}. 

However, we need to excercise caution and take into account observational bias when estimating the time lag duration on Centaurs with long orbital periods. Our integration results suggest that about half of the active Centaurs underwent $a$-jumps more than 100 years ago, but it is difficult to determine when the activity started after that. These objects were discovered and recognized as cometary bodies mostly after early 2000's, with the exception of Chiron.  After all, significant advances in telescope sensitivity and survey coverage have happened in the last few decades, meaning that we have really only had the observing capabilities to routinely detect activity on such distant and faint bodies ($V$-band magnitudes of $<21$) in the past 50 years or so, particularly compared to the length of typical Centaur orbital periods \citep[e.g.][]{Mazzotta2017}.  In addition, Centaurs as a population have also been overlooked in the past decades by all-sky surveys, with their apparent motion falling in between fast-moving near-Earth Objects (NEOs) that are the focus of surveys like Pan-STARRS \citep[e.g.][]{Chambers2016}, and slow-moving trans-Neptunian objects (TNOs) that are the focus of deep-imaging surveys like OSSOS \citep{Bannister2018}.  It is therefore entirely possible some Centaurs might have been active long before current observations flagged them as such and the decades- to centuries-long time lag preceding the activity onset is a signature of observational bias.

It is also notable that $a$-jumps only seem to activate Centaurs within a certain heliocentric distance range: all active Centaurs and JFCs have been observed at heliocentric distances of $r_h<14$~au, possibly pointing to the volatile species driving the observed activity. Several dedicated observing surveys of distant Centaurs \citep{Jewitt2009, Li2020,Cabral2019,Lilly2021a} suggest this observed lack of active objects beyond 14~au is not due to observational biases, but is instead consistent with the heliocentric range where the crystallization of AWI and/or CO$_2$ sublimation occur, implying these two processes are the main drivers of activity in the Centaur region \citep{Jewitt2009,prialnik2022}, as Centaurs mostly orbit outside the heliocentric range where H$_2$O-sublimation is possible. Indeed, the most recent JWST spectral data show presence of CO$_2$ ices on surfaces of Centaurs, and also CO$_2$ emission lines on an active Centaur 2014 OG$_{392}$ that appears stellar \citep{Licandro2023} providing observational evidence that the volatile species necessary to produce comae activity are present in this region, and that the activity could still be ongoing even if it is below our visual detection limit.

Considering the fact that recent $a$-jumps are almost exclusively present in the orbital histories of active bodies in the Centaur and JFC region, and that there are usually only short time lags between the most recent $a$-jumps and observed activity, our results suggest the $a$-jumps may act as triggers of cometary activity in Centaurs when the right conditions are met: i.e. when the $a$-jump shifts Centaur's orbit into the range of heliocentric distances where either the AWI crystallization, or CO$2$/H$2$O sublimation can take place, and, as discussed in the following sections, the orbital reshaping and/or perihelion change increases the per-orbit insolation of the Centaur. However how long does it take for the thermal wave to travel through a Centaur nucleus and take effect that remains an open question and most likely depends on the size of the object and its physical properties.

\subsection{Thermal Implications}
\label{ss.T-implications}  

Thermally, orbital re-shaping via $a$-jumps modifies the average total energy received throughout a Centaur's orbit leading to the gradual increase of surface and subsurface temperatures over a larger portion of the orbit. Thermal wave is then able to of reaching deeper layers of the nucleus, potentially producing the hottest environment the Centaurs has ever experienced, as we show in our case studies in Section~\ref{s.case_studies}. Moreover, if the $a$-jump is also accompanied by a decrease in perihelion distance, such as we see only in the few cases of the most pronounced $a$-jumps, it could lead to a temperature surge during the perihelion passages high enough to allow for sublimation of different volatile species than before. Such `double whammy' $a$-jumps then change both the energy per orbit and seasonal thermal processes on the Centaur. Perhaps the best example of this process are Centaurs LD2, 450P and also 29P. However for most Centaurs the perihelion distance value remains essentially fixed even if eccentricity and semi-major axis change after a close encounter.

We have calculated the change in the average per-orbit insolation  which was caused by the largest variation in a-e over the inspected period (for active Centaurs this change was the $a$-jump) using equation $(4)$ in \citet{Prialnik2009}. This work has shown that eccentric orbits can be modeled as circular receiving the same total energy over an
orbital period with the equivalent average circular orbit radius $a_c = a \cdot (1-e^2)$, where $a$ is semi-major axis and $e$ is the eccentricity of the original eccentric orbit. For comparison we have also used a time-averaged effective thermal radius $r_T = a \cdot (1+ \frac{1}{8}\ e^2 + \frac{21}{512}\ e^4+\mathcal{O}(e^6))$ in the calculation, which works better for estimating the energy received at distant parts of the orbit for more eccentric orbital configurations \citep{Gkotsinas2023}. We have found out the resulting pre- and post-$a$-jump average isolation change differed by only a few percent between both methods and we conclude our results are not strongly influenced by the choice of an orbit-averaging method for the type of orbits and the timescales we have investigated. 
Panel \textbf{d} on Figure~\ref{fig.ajump_examples} shows that the inactive Centaurs have typically undergone minimal and mostly negative changes in the average insolation, while active Centaurs (panel \textbf{c}) can have an increase in average insolation up to 34\% (in case of LD2). KS-test confirms that regarding the changes in insolation per orbit the null hypothesis can be rejected at the 95\% confidence level, suggesting the two groups have undergone statistically different thermo-dynamical evolution paths in the near past. 

Interestingly, Figure~\ref{fig.da_vs_q} shows that several currently inactive Centaurs have experienced $a$-jumps of similar magnitudes as active Centaurs and have perihelion distances in the range expected for active Centaurs.  Our calculations show that for the majority of these objects the orbital changes in fact did not lead to a substantial increase of per-orbit insolation, quite the opposite. There are only five notable objects that experienced both an increase in the average energy per orbit and a (gentle) decrease in both the perihelion and semi-major axis. All five Centaurs are discussed in Section~\ref{ss.Candidates}.  It is also possible that these seemingly inactive objects could have spent more time in the region and experienced prolonged periods of heating and thermal processing in the past leading to volatile depletion in the shallow subsurface reservoirs, which essentially meant they entered a period of dormancy \citep{Gkotsinas2022}. Or the differences are simply due to physical parameters such as the nucleus size, the position and depth of subsurface layers, or object's obliquity and thermal inertia, which play an important role in determining whether and when a given object will be activated \citep[e.g.][]{Prialnik2004,Guilbert2012,Davidsson2021,lisse2022}.

Centaur residence times vary by several orders of magnitude as their orbits evolve from near-Neptune parts of the Solar System towards Jupiter \citep[e.g.][]{Sarid2019,Gkotsinas2022}. They spend the majority of the time in the outskirts of the Centaur region, but once a Centaur passes the orbit of Saturn its evolution speeds up significantly and it typically takes only several orbits before it leaves the region either via the JFC pathway, or some other way. Our numerical integration results show that the magnitude of an $a$-jump is indeed dependent on the perihelion distance of the given Centaur, as can be seen on Figure~\ref{fig.da_vs_q}. 

This is because Centaurs with smaller perihelia have a higher chance to undergo a series of close encounters with Saturn or Jupiter that drive the $a$-jumps. This relationship between the $a$-jump magnitude and perihelion distance also indicates that the more distant Centaurs need a smaller impulse to be activated, and this is very likely because these bodies have just emerged from the `freezer', and have undergone very little, if any, thermal processing yet. These distant Centaurs most likely still contain ice deposits on the surface \citep{Licandro2023} or in shallow subsurface layers \citep[e.g.][]{Barucci2002}, and are able to undergo phase transition even with small temperature change, which appears to be the case with 167P and a handful of other active Centaur with high perihelia and mild $a$-jumps ($\Delta a <0.3$~au).  With these objects, their $a$-jumps occur at the end of a drift period, during which semi-major axis and eccentricity slowly decrease over several millennia. Mild $a$-jumps are then produced by distant close encounters with Saturn, typically within several Hill radii. We speculate that once these ``drifters'' reach sufficiently small semi-major axes, more pronounced close encounters will start occuring, propelling them onto the fast evolutionary track towards JFC orbits. Apart from 167P, there are two other active drifter Centaurs with similar orbital history: 2003 QD$_{112}$ and (248835) 2006 SX$_{368}$.

Our results are in line with thermal modeling work by \citet{Gkotsinas2022}, who show that Centaur progenitors of the JFCs undergo multiple periodic heating episodes during their transition through Centaur region that warm up consequently deeper and deeper layers of the nuclei. Centaurs gradually lose volatiles during every episode until the reservoirs of the thermally pristine or unprocessed material are either deep enough that even at perihelion distances typical for JFC region the thermal wave doesn't reach them, or smaller Centaur nuclei will be thermally processed all the way through.  In fact, the time spent in transient orbits is critical - \citet{Gkotsinas2022} have shown that JFCs spending long periods of time on transient orbits (i.e., with longer lifetime) are more susceptible to be heated up at greater depths and are prone to losing all free condensed hypervolatiles down to their core. This would explain why some inactive Centaurs have orbits very similar to active objects - these bodies are either completely processed and dormant, or on the other hand they could be `dynamically younger' and could still have pockets of volatiles in the deeper layers with the possibility of reactivation if a future $a$-jump can reshape the orbit again and thermal wave could propagate deeper. 

Our case study of 167P (Section~\ref{s.case_studies}) further underscores that the thermal processing the nucleus had undergone and the depth of its ice deposits might play an important role - dynamically new objects with volatile pockets in shallow underground layers can be reached by thermal wave that penetrates only several cm- to m-deep and is triggered by a mild a-jump. In contrast, bodies that have undergone multiple periods of activity and have their upper layers depleted of volatiles likely need a major a-jump that allows for more pronounced heating and a subsequent thermal wave reaching deeper parts of the nucleus' interior.

Does this mean that every new Centaur has the potential to be active under certain circumstances? It is likely, since they have formed beyond snowlines of several volatile species, and even though primordial CO ice deposits have most likely already sublimated since their formation time \citep{Lisse2021, Licandro2023}, the majority of Centaurs could be expected to contain volatile pockets hidden deeper in the nucleus that have remained untouched until present day. As the Centaurs are making their way through the region by the means of planetary perturbations the sudden change in thermal conditions induced by $a$-jumps could eventually reach the sublimation temperatures for different volatile species, or the threshold at which AWI crystallizes, and set in motion a chain of events leading to an outburst. This way, a Centaur will transition between different activity drivers throughout its dynamical evolution. Furthermore, depending on the type of volatiles present and the size of the deposits, Centaurs activated this way could sustain gas and dust production for months, or even for several orbits depending on the speed at which the thermal wave propagates through the body until the activity-driving material is depleted. For example, \citet{Lilly2021a} proposed that the propagation of the AWI crystallization front and sudden pressure from escaped volatile molecules could burst out and create an opening on the surface, or a landslide and expose the volatiles hidden under several skin depths, which could sublimate and sustain activity. Such regional events and large pits have been observed in situ on comet 67P by the Rosetta spacecraft \citep{Steckloff2018,Jindal2022}.

\subsection{Identifying future active Centaur candidates}
\label{ss.Candidates} 

We use our $a$-jump hypothesis to identify Centaurs that have the potential to become candidates for outbursts in the near future. \citet{Jewitt2009,Li2020,Lilly2021a} have shown that the Centaur activity is associated with orbits that have perihelion distance less than $\sim14$~au, because the vast majority of active Centaurs were at the heliocentric distances $r_h<14$~au when they were detected, with the sole exception of Chiron. However, Chiron is an unusual object, the second largest Centaur known ($r\sim120$~km), and apart from its activity, there is evidence of a ring system surrounding it \citep{Ortiz2015}. 

Based on our results, a good future outburst candidate would be: a) a seemingly inactive object from the low-perihelion group ($q<10$~au), b) one that is currently at heliocentric distance less than $r_h<14$~au, and c) one that has experienced an $a$-jump deeper than 0.5~au in the last 200 years which yielded a positive change in the average per orbit insolation $>$2\%. Another favorable condition for possible observable activity would be if the Centaur is also nearing its next perihelion passage. A particularly interesting indicator of past activity could also be photometric surface colors, as active Centaurs have almost exclusively photometrically neutral surfaces akin to JFCs, which were proposed to be the results of blanketing by ejecta \citep{Jewitt2015}.

There are currently only 12 Centaurs in the MPC database that fulfill the first three criteria, but only three of them are approaching perihelion in the next 15 or so years:  534251 (2014 SW223), a small , D$\sim$10~km Centaur that experienced a large, 1.7 au $a$-jump 360 years ago, (31824) Elatus, a medium size Centaur with a diameter of $r\sim28$~km that experience a 1~au $a$-jump 10 years ago; and (32532) Thereus, a large Centaur with $r\sim43$~km, that underwent a 0.9~au $a$-jump 100 years ago. All three objects should be considered high-priority targets for observational monitoring to search for activity. From the three only Thereus and Elatus have measured photometric colors. Thereus falls into the gray group \citep{Tegler2003,DeMeo} which possibly suggests it could have undergone activity periods in the past, while Elatus lies on the outskirts of the red group hinting on primordial red surface \citep{Tegler2003,Peixinho2003}. The three Centaurs are currently at heliocentric distances of 10.3 au, 15.5~au and 11.6~au, with apparent brightness of $V=23.3$~mag$, V=22.6$~mag and $V=20.2$~mag, respectively. The remaining nine Centaurs from the group (2013 CJ$_{118}$, 2015 KH$_{172}$, 2013 AS$_{105}$, 2014 ON$_{6}$, 2008 RG$_{167}$, 2004 VP$_{112}$, 2014 LR$_{14}$, 2010 LJ$_{109}$, 2001 XA$_{255}$) are currently moving away from perihelion, fading rapidly past V$\sim$25, and as they have orbital periods comparable to human lifespan, it will be up to the next generation of astronomers to search them for possible future activity. 

\section{Conclusions}
\label{s.Conclusions} 

We present the results from an analysis of the individual dynamical histories of Centaurs and high-perihelion JFCs aimed at identifying the triggers of cometary activity. We have applied a thermal model to two end-member cases of known active Centaurs to investigate the thermal evolution and onset of activity coupled with their orbital histories. Our results show that:

\begin{enumerate}
    \item All known active Centaurs and JFCs studied in this work have undergone a dramatic and rapid orbit reshaping in the near past, characterized by a sudden drop in semi-major axis and eccentricity. These $a$-jumps occur on timescales less than a decade, sometimes just months, and are able to change the original semi-major axis by several leading to a significant increase in the average per-orbit insolation. We suggest $a$-jumps could be potential triggers of cometary activity on some Centaurs, and act by significantly shortening orbital periods and placing the objects into warmer environments, producing heatwaves reaching down into thermally unprocessed layers of the nuclei. A typical Centaur likely experiences several $a$-jumps during its transition into a JFC once it passes the orbit of Saturn.

    \item  We applied a thermal model to the orbital histories of two active Centaurs, 167P/CINEOS and P/2019 LD2 (ATLAS), orbiting on opposite ends of the heliocentric distance range within which active Centaurs are found in order to investigate how $a$-jumps influence the thermal environments inside their nuclei. We conclude that $a$-jumps led to a temperature increases of more than 2K and 20K on 167P and LD2, respectively, while thermal waves propagated deeper into each nucleus by more than 2~m. Such abrupt changes in internal temperatures seem likely to have set off phase changes in ices present in the structure of the nucleus. Temperatures changes reached inside 167P near the sub-solar point would have been too low to produce activity via AWI crystallization, but too high for CO and CO$_2$ sublimation. The activity driver for 167P and its connection with the more recent $a$-jump in the 1800s is poorly understood. Surface and near-sub-surface CO$_2$ ice deposits at higher latitudes could be the driver of activity, triggered by the slow inward drift observed for its orbit. LD2 on the other hand possesses a much more pronounced temperature change after its $a$-jump, where gradually temperatures reached high enough values for the onset of vigorous H$_2$O sublimation. Our modeling results suggest that $a$-jumps were the activity triggers for our two test Centaurs.

    \item Based on the identification of $a$-jumps in the orbital histories of several known currently inactive Centaurs with perihelion distances compatible with either AWI crystallization or CO$_2$ sublimation, we predict that these objects may be observed to exhibit activity in the near future, and strongly encourage observational monitoring of three particular targets by the community to search for that activity.
    
\end{enumerate}

\section{Acknowledgements}
\label{s.Acknowledgements}
\begin{acknowledgements}

This work was supported by NSF AST Grant \#1910275. We additionally acknowledge support from the Florida Space Research Initiative program.

This research has made use of data and/or services provided by the International Astronomical Union's Minor Planet Center. 

We thank the two anonymous reviewers for helpful comments that pointed out details of thermal processing of cometary nuclei and led to significant improvement of the manuscript.
\end{acknowledgements}

\bibliography{Centaurs}{}
\bibliographystyle{aasjournal}


% T A B L E S -------------------------------------------------------
\startlongtable
\begin{deluxetable}{lccccccccc}
\tablewidth{370pt} 
\tabletypesize{\footnotesize}
\tablecolumns{10}
\tablecaption{Orbital and $a$-jump parameters of investigated active Centaurs and high-perihelion JFCs at the epoch in the beginning of the integration (last column).}
\tablehead{
\colhead{Name} & 
\colhead{q$^a$} & 
\colhead{a$^b$} & 
\colhead{e$^c$} & 
\colhead{i$^d$} &
\colhead{P$^e$} & 
\colhead{$T_J^f$} & 
\colhead{$\Delta_{a}^g$} & 
\colhead{t$_{0}^h$} & 
\colhead{Epoch$^i$}
}
\startdata
\textbf{Centaurs}	&		    &		    &		    &		    &		    &		    &		    &		&  \\
P/2019 LD2 (ATLAS)	&	4.58	&	5.29	&	0.134	&	11.6	&	12.2	&	2.94	&	-3.0	&	0	&	2459098.5	\\
P/2010 C1 (Scotti)	&	5.24	&	7.06	&	0.259	&	9.1	&	18.8	&	2.96	&	-3.0	&	-143	&	2455264.5	\\
P/2020 W1 (Rankin)	&	5.29	&	7.19	&	0.265	&	10.8	&	19.3	&	2.95	&	-0.4	&	-38	&	2459161.5	\\
C/2001 M10 (NEAT)	&	5.30	&	26.65	&	0.801	&	28.1	&	137.7	&	2.59	&	-0.8	&	-17	&	2452216.5	\\
P/2011 C2 (Gibbs)	&	5.39	&	7.36	&	0.268	&	10.9	&	20.0	&	2.96	&	-2.2	&	-132	&	2455840.5	\\
C/2020 Q2 (PANSTARRS)	&	5.40	&	10.91	&	0.505	&	3.3	&	36.1	&	2.97	&	-0.6	&	0	&	2459071.5	\\
P/2008 CL94 (Lemmon)	&	5.43	&	6.17	&	0.119	&	8.3	&	15.3	&	2.98	&	-1.4	&	-96	&	2454769.5	\\
P/2017 P1 (PANSTARRS)	&	5.44	&	7.86	&	0.308	&	7.7	&	22.1	&	2.98	&	-0.4	&	-62	&	2457994.5	\\
39P/Oterma	&	5.46	&	7.21	&	0.243	&	2.0	&	19.4	&	3.00	&	-2.8	&	-80	&	2456910.5	\\
C/2015 V4 (PANSTARRS)	&	5.46	&	18.57	&	0.706	&	60.8	&	79.9	&	1.59	&	-0.4	&	-151	&	2457465.5	\\
P/2004 A1 (LONEOS)	&	5.46	&	7.89	&	0.308	&	10.6	&	22.2	&	2.96	&	-5.2	&	-28	&	2453385.5	\\
C/2007 S2 (Lemmon)	&	5.56	&	12.55	&	0.557	&	16.9	&	44.4	&	2.88	&	-1.1	&	-92	&	2454610.5	\\
C/2015 D2 (PANSTARRS)	&	5.61	&	12.98	&	0.568	&	31.8	&	46.8	&	2.61	&	-0.2	&	0	&	2457104.5	\\
C/2014 F3 (Sheppard-Trujillo)	&	5.72	&	16.07	&	0.644	&	6.5	&	64.5	&	3.00	&	-0.7	&	0	&	2456918.5	\\
29P/Schwassmann-Wachmann1	&	5.73	&	5.99	&	0.045	&	9.4	&	14.7	&	2.98	&	-0.5	&	-45	&	2459945.5	\\
174P/Echeclus	&	5.82	&	10.67	&	0.455	&	4.3	&	34.9	&	3.03	&	-0.6	&	-180	&	2460000.5	\\
P/2015 M2 (PANSTARRS)	&	5.91	&	7.20	&	0.179	&	4.0	&	19.3	&	3.03	&	-0.4	&	-8	&	2457418.5	\\
C/2013 P4 (PANSTARRS)	&	5.97	&	14.77	&	0.596	&	4.3	&	56.8	&	3.05	&	0.3	&	-174	&	2456851.5	\\
P/2015 B1 (PANSTARRS)	&	5.98	&	9.67	&	0.382	&	18.0	&	30.1	&	2.93	&	-0.2	&	0	&	2457101.5	\\
2020 MK4	&	6.03	&	6.15	&	0.02	&	6.7	&	15.2	&	3.01	&	-0.8	&	-190	&	2459800.5	\\
P/2010 H5 (Scotti)	&	6.03	&	7.14	&	0.156	&	14.1	&	19.1	&	2.97	&	-0.7	&	-73	&	2455337.5	\\
C/2011 P2 (PANSTARRS)	&	6.15	&	9.76	&	0.37	&	9.0	&	30.5	&	3.05	&	-0.6	&	-560	&	2456851.5	\\
(523676) 2013 UL10	&	6.20	&	9.95	&	0.377	&	19.2	&	31.2	&	2.94	&	-0.2	&	-32	&	2459396.5	\\
P/2005 T3 (Read)	&	6.20	&	7.51	&	0.174	&	6.3	&	20.6	&	3.04	&	-0.6	&	-114	&	2453667.5	\\
P/2020 V3 (PANSTARRS)	&	6.23	&	8.36	&	0.255	&	23.0	&	24.2	&	2.88	&	-0.4	&	-340	&	2459176.5	\\
C/2019 A5 (PANSTARRS)	&	6.32	&	21.66	&	0.708	&	67.5	&	100.7	&	1.34	&	-0.3	&	0	&	2458701.5	\\
P/2005 S2 (Skiff)	&	6.40	&	7.97	&	0.197	&	3.1	&	22.5	&	3.08	&	-0.9	&	-490	&	2452114.5	\\
165P/LINEAR	&	6.83	&	18.07	&	0.622	&	15.9	&	76.7	&	3.09	&	-0.6	&	-220	&	2452114.5	\\
P/2011 S1 (Gibbs)	&	6.89	&	8.65	&	0.203	&	2.7	&	25.5	&	3.12	&	-2.8	&	-799	&	2457635.5	\\
C/2016 Q4 (Kowalski)	&	7.08	&	16.79	&	0.578	&	7.3	&	68.8	&	3.22	&	-0.7	&	-132	&	2457635.5	\\
2003 QD$_{112}$	&	7.92	&	18.98	&	0.583	&	14.5	&	82.6	&	3.28	&	-0.7	&	-250	&	2460000.5	\\
95P/Chiron	&	8.51	&	13.71	&	0.379	&	6.9	&	50.7	&	3.36	&	-0.7	&	-850	&	2460000.5	\\
166P/NEAT	&	8.56	&	13.88	&	0.383	&	15.4	&	51.7	&	3.28	&	1.2	&	-2240	&	2452819.5	\\
C/2013 C2 (Tenagra)	&	9.13	&	16.05	&	0.431	&	21.3	&	64.3	&	3.28	&	-0.1	&	-70	&	2457165.5	\\
C/2015 T5 (Sheppard-Tholen)	&	9.34	&	27.96	&	0.666	&	11.0	&	147.9	&	3.58	&	-0.6	&	-4	&	2457376.5	\\
C/2012 Q1 (Kowalski)	&	9.48	&	26.12	&	0.637	&	45.2	&	133.6	&	2.63	&	-0.4	&	-5	&	2456322.5	\\
P/2014 OG$_{392}$	&	9.95	&	12.15	&	0.181	&	9.0	&	42.3	&	3.40	&	-0.4	&	-163	&	2458289.5	\\
167P/CINEOS	&	11.78	&	16.14	&	0.27	&	19.1	&	64.8	&	3.53	&	-0.2	&	-146	&	2453096.5	\\
(248835) 2006 SX$_{368}$	&	11.95	&	22.01	&	0.457	&	36.3	&	103.2	&	3.18	&	-0.3	&	-110	&	2455545.5	\\
\textbf{JFCs}	        &		    &		    &		    &		    &		    &		    &		    &		&   \\
C/2018 P5	(PANSTARRS)	&	4.58	&	12.75	&	0.641	&	 7.3 	&	45.5	&	2.792	&	-2.0	&	-4	&	2458531.5	\\
P/2015 D6	(Kowalski)	&	4.58	&	9.15	&	0.500	&	 20.4 	&	27.7	&	2.722	&	3.0	&	-140	&	2457104.5	\\
158P	Kowal-LINEAR	&	4.58	&	4.72	&	0.030	&	 7.9 	&	10.3	&	2.988	&	-2.5	&	-75	&	2455827.5	\\
P/2014 O3	(PANSTARRS)	&	4.64	&	7.55	&	0.385	&	 7.8 	&	20.7	&	2.892	&	-0.6	&	-110	&	2456855.5	\\
P/2018 V5	(Trujillo-Sheppard)	&	4.71	&	8.97	&	0.475	&	 10.6 	&	26.8	&	2.852	&	-1.1	&	-53	&	2458331.5	\\
P/404P	Bressi	&	4.76	&	5.21	&	0.085	&	 9.7 	&	11.9	&	2.964	&	-1.0	&	0	&	2458048.5	\\
P/2016 A3	(PANSTARRS)	&	4.79	&	7.73	&	0.380	&	 8.6 	&	21.4	&	2.903	&	-1.3	&	-49	&	2457890.5	\\
P/2010 H4	(Scotti)	&	4.82	&	6.62	&	0.272	&	 2.3 	&	17.0	&	2.955	&	-1.2	&	-170	&	2455321.5	\\
P/2012 T2	(PANSTARRS)	&	4.82	&	5.74	&	0.160	&	 12.6 	&	13.7	&	2.930	&	-3.0	&	-12	&	2455793.5	\\
C/2008 E1	(Catalina)	&	4.83	&	10.68	&	0.548	&	 35.0 	&	34.9	&	2.450	&	0.7	&	-137	&	2454811.5	\\
P/2015 PD$_{229}$	(Cameron-ISON)	&	4.83	&	7.18	&	0.327	&	 2.0 	&	19.2	&	2.944	&	-0.6	&	-900	&	2457316.5	\\
P/2010 U1	(Boattini)	&	4.90	&	6.63	&	0.261	&	 8.2 	&	17.1	&	2.942	&	-1.5	&	-5	&	2455524.5	\\
P/2011 P1	(McNaught)	&	4.96	&	7.74	&	0.360	&	 6.2 	&	21.5	&	2.935	&	-2.2	&	-250	&	2455793.5	\\
P/2011 JB15	(Spacewatch-Boattini)	&	5.02	&	7.37	&	0.319	&	 19.1 	&	20.0	&	2.837	&	-0.9	&	-35	&	2455721.5	\\
P/2019 V2	(Groeller)	&	5.02	&	7.53	&	0.333	&	 11.8 	&	20.6	&	2.912	&	-7.0	&	-650	&	2459056.5	\\
P/377P	Scotti	&	5.04	&	6.72	&	0.251	&	 9.0 	&	17.5	&	2.947	&	-4.0	&	-250	&	2456842.5	\\
P/2020 V4	(Rankin)	&	5.15	&	9.34	&	0.449	&	 14.2 	&	28.5	&	2.878	&	-0.9	&	-110	&	2459056.5	\\
\enddata
\tablenotetext{*}{\textbf{Notes.}\\
$^a$ Perihelion distance, AU.\\
$^b$ Semimajor axis, AU.\\
$^c$ Orbital eccentricity\\
$^d$ Inclination, degrees.\\
$^e$ Current orbital period, years.\\
$^f$ Tisserand parameter with respect to Jupiter.\\
$^g$ Change in semi-major axis during most recent a-jump, AU.\\
$^h$ End of the most recent a-jump, years from 2022. Zero means the a-jump is still in progress. \\
$^i$ Orbital epoch of the covariance matrix at the start of numerical integration, Julian Day.
}
\label{tab.Orbparameters}
\end{deluxetable}

% F I G U R E S -------------------------------------------------------
\newpage

\end{document}